\documentclass[aip,jcp,reprint,showpacs,numerical]{revtex4-1}

\usepackage{amsmath}

\newcommand{\V}[1]{\mathbf{#1}}
\newcommand{\bracket}[1]{\left(#1\right)}

\newcommand{\eq}[1]{$\mathrm{Eq.}$~\eqref{#1}}
\newcommand{\av}[1]{\left\langle #1 \right\rangle}

\makeindex

\begin{document}

\title{From Reaction-Diffusion Systems to Confined Brownian Motion}
\author{S. Martens}
\email{steffen.martens@tu-berlin.de}
\affiliation{Institut f\"ur Theoretische Physik, Technische Universit\"at Berlin, Hardenbergstra\ss e 36, 10623 Berlin, Germany}
\maketitle

\noindent Propagation of traveling wave (TW) patterns, including traveling fronts, solitary excitation pulses, and scroll waves, plays an important role in various technological and biophysical processes such as catalysis \cite{Bar1992,*Corma1997,*Santamaria2016}, chemical computing \cite{Steinbock1996,*Totz2015b}, neural information processing \cite{koch2000}, atrial arrhythmia \cite{Fenton2011}, pattern formation in the cell cortex \cite{Bois2011,*bement2015}. Often, the excitable medium supporting wave propagation exhibits an irregular shape and/or is limited in size, leading to complex wave phenomena \cite{Toth1994,*vanag2001pattern,*Totz2015,*biktasheva_drift_2015}. In particular, the analytic treatment of these wave phenomena is notoriously difficult due to the spatial modulation of the domain's boundary leading to space-dependent no-flux boundary conditions (NFBCs) on the surface. 

Recently \cite{Martens2015,Ziepke2016arxiv}, we have provided a first systematic treatment by applying asymptotic perturbation analysis in a geometric parameter \cite{Note1}, determining the domain's cross section changing rate. This led to an approximate description 
that involves a reduction of dimensionality; the $3$D reaction-diffusion equation (RDE) with spatially dependent NFBCs on the reactants reduces to a $1$D reaction-diffusion-advection equation (RDAE) for the leading order
\begin{align} \label{eq:RDA}
\partial_t u(\V{r},t)=\,D_u\partial_x^2 u + D_u\,Q'\bracket{x}/Q\bracket{x}\partial_x u+R(u).
\end{align}
For simplicity, here, we consider the single component case in which $u(\V{r},t)$ corresponds to the concentration of species $u$ at position $\V{r}=(x,y,z)^T$ and at time $t$, $D_u$ denotes the molecular diffusion constant of $u$, and $R(u)$ represents the local, nonlinear reaction kinetics. Further, we presume that the domain's cross section $Q(x)$ is $L$-periodic in $x$-direction. In contrast to the conventional approach used for diffusion problems in confined domains \cite{Note2}, \eq{eq:RDA} is consistent with the eikonal equation \cite{Keener1986,*Dierckx2011}. Considering exemplarily a chemical front propagating from left to right, $\partial_x u < 0$, the boundary-induced advection field $\V{v}=- D_u\,Q'/Q\,\V{e}_x$ guaranties that fronts become decelerated where the channel expands, $Q'>0$, and accelerated if the channel contracts, $Q'<0$, respectively. 

Interestingly, in the high-diffusive limit we found that the propagation velocities of traveling fronts \cite{Martens2015} and solitary excitation pulses \cite{Ziepke2016arxiv} in spatially modulated domains saturate at a value depending solely on the geometry of the domain. In the latter, the intrinsic width of the TW, $l \propto \sqrt{D_u}$, is much larger than the period of the modulation $L$. We argued heuristically that diffusion of reactants in propagation direction under spatially confined conditions is the predominant process for wave propagation and boundary interactions play a subordinate role in this limit. Consequently, the problem of wave propagation might be approximated well by a quasi $1$D description introducing an effective diffusion constant $D_\mathrm{eff}$ to the RDE; yielding
\begin{align}  \label{Eq:effRDE}
\partial_t u(\V{r},t) =\, D_\mathrm{eff} \partial_x^2 u + R(u).
\end{align} 
Experimental \cite{Verkman2002,*Cohen2006} and theoretical studies \cite{Dagdug2011,*Martens2013,*Kalinay2014} on particle transport in micro-domains with obstacles \cite{Dagdug2012a,*Martens2012} and/or small openings revealed non-intuitive features like a significant suppression of particle diffusivity -- also called confined Brownian motion. Numerous research activities led to the development of an approximate description of the transport problem -- the \textit{Fick-Jacobs approach} \cite{Burada2008,*Burada2009_CPC}. The latter provides a powerful tool to capture many properties of particle transport and predicts that the effective diffusion coefficient in channel direction $D_\mathrm{eff}$ is solely determined by the cross section $Q(x)$. Thereby, $D_\mathrm{eff}$ can be calculated according to the Lifson-Jackson formula\cite{Lifson1962}
\begin{align}
\label{Eq:LifsonJackson}
D_{\mathrm{eff}}/D_u=\,\left[\langle Q(x)\rangle_{L}\langle Q^{-1}(x)\rangle_{L} \right]^{-1},
\end{align}
where $\av{\bullet}_{L}=L^{-1}\int_{0}^{L}\bullet\,\mathrm{d}x$ denotes the average mean over one period of the modulation. We emphasize that both the Fick-Jacobs equation and the Lifson-Jackson formula represent the leading order solution in an asymptotic perturbation analysis 
of the associated stationary Smoluchowski equation and $B$-field \cite{Dorfman2014}, respectively. Importantly, our heuristic argumentation resulting in \eq{Eq:effRDE} has been already confirmed numerically \cite{Martens2015,Ziepke2016arxiv}. 

In this note, we demonstrate how one derives the quasi $1$D RDE with an effective diffusion coefficient $D_\mathrm{eff}$ given by \eq{Eq:LifsonJackson} from the RDAE, \eq{eq:RDA}.
Rescaling space by the diffusion coefficient, $x \to \sqrt{D_u} x$, yields
\begin{align} \label{Eq:AppRDA}
\partial_t u(x,t)=&\,\partial_x^2 u+ Q'\bracket{x/\alpha}/Q\bracket{x/\alpha} \partial_x u+R(u),
\end{align}
where $\alpha=\,L/\sqrt{D_u} \propto L/l$ is proportional to the ratio of modulation period $L$ to intrinsic width $l$. We focus on the limit $\alpha \to 0$ in which the advective velocity field $\V{v}=-Q'(x/\alpha)/Q(x/\alpha)\V{e}_x$ changes rapidly, periodically in space and apply homogenization theory \cite{Keener2000,*xin_front_2000} to \eq{Eq:AppRDA}. Importantly, we suppose that the variations on the microscopic ($\alpha$-) and macroscopic ($x$-) scales can be described by two variables, $\chi=x$ and $\sigma=x/\alpha$, both being treated as independent quantities. By introducing these two independent scales and transforming the spatial derivative accordingly, $\partial_x = \partial_{\chi} + 1/\alpha \partial_\sigma$, \eq{Eq:AppRDA} becomes
\begin{align} \label{Eq:AppRDA2}
\begin{split}
 \alpha^2\,\partial_t u(\chi,\sigma,t)&=\,\bracket{\alpha^2 \partial_{\chi}^2 + 2\alpha \partial_{\chi}\partial_\sigma + \partial_\sigma^2} u+ \\ &\!\partial_\sigma Q(\sigma)/Q(\sigma)\bracket{\alpha \partial_\chi+\partial_\sigma}u+\alpha^2\,R(u).
\end{split}
\end{align}
Next, we presume that any solution to \eq{Eq:AppRDA2} can be expanded in a series in $\alpha$
\begin{align} \label{Eq:u_series}
 u(\chi,\sigma,t)\!=\!u_0(\chi,t)\!+\!\alpha\, u_1(\chi,\sigma,t)\!+\!\alpha^2 u_2(\chi,\sigma,t)\!+\!\ldots
\end{align}
In such a representation for $u$, the leading order term $u_0(\chi,t)$ represents the average, or mean field, behavior of $u$ on the macroscopic scale $\chi$. Hence, all higher terms have zero mean in $\sigma$, $\av{u_i}=0,\,\forall\,i \geq 1$, with $\av{\bullet}=\int_0^1 \bullet\, \mathrm{d}\sigma$, and $u_0$ can be assumed to be independent of $\sigma$. Moreover, we claim that all $u_i$ are $1$-periodic, $u_i(\chi,1,t)=u_i(\chi,0,t),\,\forall\,i\geq 1$, because the advection term $\propto Q'(\sigma)/Q(\sigma)$ changes with period $1$. Inserting \eq{Eq:u_series} into \eq{Eq:AppRDA2} leads to a hierarchic set of coupled partial differential equations
\begin{subequations}
 \begin{alignat}{2}
 {}\mathcal{O}(\alpha^0):&\,& 0&=\,\partial_\sigma\bracket{Q(\sigma)\partial_\sigma u_0} \label{Eq:order0}\\
 {}\mathcal{O}(\alpha^1):&\,& 0&=\,\partial_\sigma\bracket{Q(\sigma)\left\lbrace \partial_\chi u_0 + \partial_\sigma u_1\right\rbrace} \label{Eq:order1}\\
 \mathcal{O}(\alpha^2):&\,& \partial_t u_0&=\,\partial_\sigma\bracket{Q(\sigma)\partial_\sigma u_2}/Q(\sigma)+2\partial_\chi\partial_\sigma u_1  \notag \\
 &&&+\partial_\chi^2 u_0+Q'(\sigma)/Q(\sigma)\partial_\chi u_1+R(u_0). \label{Eq:order2}
 \end{alignat}
\end{subequations}
Obviously, the zeroth order equation \eq{Eq:order0} is always fulfilled for any function $u_0$ being independent of $\sigma$. From \eq{Eq:order1}, one gets $\partial_\chi u_0 + \partial_\sigma u_1 = \,K(\chi,t)/Q(\sigma)$ where the unknown function $K(\chi,t)$ is determined by the periodicity requirement in $\sigma$ for $u_1(\chi,\sigma,t)$. Integrating the equation for $K(\chi,t)$ over $\sigma$, yields $\partial_\sigma u_1=\,\partial_\chi u_0\, \left[1/(Q(\sigma) \av{Q^{-1}})-1\right]$. Plugging the result for $\partial_\sigma u_1$ into \eq{Eq:order2} and multiplying the latter by $Q(\sigma)$ results in
\begin{align*}
 Q(\sigma)\partial_t u_0={}&\,\partial_\sigma\bracket{Q(\sigma)\partial_\sigma u_2}+ \partial_\chi^2 u_0 \left[2\,\av{Q^{-1}}^{-1}-Q(\sigma)\right]\\
 &\,+Q'(\sigma)\partial_\chi u_1+Q(\sigma)\,R(u_0). 
\end{align*}
By integrating the last equation over $\sigma$ and taking the periodicity of $u_1$ and $u_2$ into account, one obtains the homogenized RDE for the leading order $u_0$
\begin{align}
 \partial_t u_0(\chi,t)=&\left[\av{Q}\av{Q^{-1}}\right]^{-1}\partial_\chi^2 u_0(\chi,t)+R(u_0). \notag 
 \intertext{We get our final result}
 \partial_t u_0(x,t)=&D_u\left[\av{Q}_{L} \av{Q^{-1}}_{L}\right]^{-1}\partial_x^2 u_0+R(u_0), \label{Eq:RDE_Deff}
\end{align}
by scaling the macroscopic length scale $\chi$ back to the real domain, $\sigma \to x \sqrt{D_u}/L$. Comparing \eq{Eq:effRDE} with \eq{Eq:RDE_Deff}, one identifies the effective diffusion coefficient by $D_\mathrm{eff} = D_u / \bracket{\av{Q}_{L} \av{Q^{-1}}_{L}}$ which is the main result of this note. 

We demonstrated for the first time that one can derive an expression for the effective diffusion coefficient, equal to the Lifson-Jackson formula, using a subsequent homogenization of the $1$D reaction-diffusion-advection equation. The latter has been derived by applying asymptotic perturbation analysis to the underlying $3$D reaction-diffusion equation with spatially dependent no-flux boundary conditions and incorporates the effects of boundary interactions on the reactants via a boundary-induced advection term. We stress that both quasi one-dimensional descriptions, the reaction-diffusion-advection equation and the Fick-Jacobs equation, are solely accurate for weakly modulated confinements.

I thank Alexander Ziepke for fruitful discussions and acknowledge financial support from the German Science Foundation DFG through SFB 910.

\end{document}